\documentclass[12pt]{article}
\usepackage{a4wide}
\usepackage{epsfig}
\usepackage{amsmath}

\newcommand{\half}{{\textstyle\frac{1}{2}}}

\newlength{\absize}
\catcode`@=11
\def\citer{\@ifnextchar [{\@tempswatrue\@citexr}{\@tempswafalse\@citexr[]}}

\def\@citexr[#1]#2{\if@filesw\immediate
  \write\@auxout{\string\citation{#2}}\fi
  \def\@citea{}\@cite{\@for\@citeb:=#2\do
    {\@citea\def\@citea{--\penalty\@m}\@ifundefined
       {b@\@citeb}{{\bf ?}\@warning
       {Citation `\@citeb' on page \thepage \space undefined}}%
\hbox{\csname b@\@citeb\endcsname}}}{#1}}
\catcode`@=12

\begin{document}
  \thispagestyle{empty}
  \pagestyle{empty}
  \renewcommand{\thefootnote}{\fnsymbol{footnote}}
\newpage\normalsize
    \pagestyle{plain}
    \setlength{\baselineskip}{4ex}\par
    \setcounter{footnote}{0}
    \renewcommand{\thefootnote}{\arabic{footnote}}
\newcommand{\preprint}[1]{%
  \begin{flushright}
    \setlength{\baselineskip}{3ex} #1
  \end{flushright}}
\renewcommand{\title}[1]{%
  \begin{center}
    \LARGE #1
  \end{center}\par}
\renewcommand{\author}[1]{%
  \vspace{2ex}
  {\Large
   \begin{center}
     \setlength{\baselineskip}{3ex} #1 \par
   \end{center}}}
\renewcommand{\thanks}[1]{\footnote{#1}}
\begin{flushright}
{hep-th/0102014} \\[2pt]
February 2001
\end{flushright}
\vskip 0.5cm

\begin{center}
{\large \bf Perturbative Aspects of $q$-Deformed Dynamics}
\end{center}
\vspace{1cm}
\begin{center}
Jian-zu Zhang$^{a,b,c,\S}$ and P. Osland$^{b,*}$
\end{center}
\vspace{1cm}
\begin{center}
$^a$ Department of Physics,
University of Kaiserslautern, PO Box 3049, D-67653  Kaiserslautern, Germany \\
$^b$ Department of Physics,
University of Bergen, N-5007 Bergen, Norway \\
$^c$ Institute for Theoretical Physics, Box 316,
East China University of Science and Technology,
Shanghai 200237, P. R. China
\end{center}
\vspace{1cm}

\begin{abstract}
Within the framework of the $q$-deformed Heisenberg algebra a dynamical
equation of $q$-deformed quantum mechanics is discussed. 
The perturbative aspects of the $q$-deformed Schr\"odinger equation 
are analyzed.
General representations of the additional momentum-dependent interaction
originating from the $q$-deformed effects are presented in two approaches. 
As examples, such additional interactions related to 
the harmonic-oscillator potential and the Morse potential are demonstrated.
\end{abstract}

\begin{flushleft}
${^\S}$ E-mail address: jzzhangw@online.sh.cn\\
${^*}$ E-mail address: Per.Osland@fi.uib.no 
\end{flushleft}
\clearpage

Recently $q$-deformed quantum mechanics has attracted much attention
as a possible modification of the ordinary quantum mechanics at 
short distances. 
According to present tests of quantum electrodynamics, quantum theories 
based on Heisenberg's commutation relation are correct at least 
down to $10^{-18}$~cm. 
The question arises whether there is a possible generalization of 
Heisenberg's commutation relation at shorter distances. 
In searching for such a possibility considerations 
of the space structure are a useful guide. 
If the space structure at such short distances exhibits 
a non-commutative property, and thus is governed by a quantum group symmetry, 
it has been shown that $q$-deformed quantum mechanics is a 
possible pre-quantum theory at short distances. 
In the literature different frameworks of $q$-deformed quantum mechanics 
were established \citer{Schwenk:1992sq,Cerchiai:1999eg}. 

The framework of the $q$-deformed Heisenberg algebra developed 
in Refs.~\cite{Hebecker:1994eb,Fichtmuller:1996dt} shows 
clear physical content: its relation to the corresponding
$q$-deformed boson commutation relations and the limiting process of
the $q$-deformed harmonic oscillator to the undeformed one are clear. 
In this framework the $q$-deformed uncertainty relation shows 
an essential deviation from that of Heisenberg \cite{JZZ99}: 
the ordinary minimal uncertainty relation is undercut.
A non-perturbative feature of the $q$-deformed 
Schr\"odinger equation is that the energy
spectrum exhibits an exponential structure \cite{Lorek:1995du,Fichtmuller:1996dt,Zhang:2000qj}. 
The pattern of quark and lepton masses is qualitatively explained by such a 
$q$-deformed  exponential spectrum \cite{Zhang:2000qj}. 

In this paper we discuss {\it perturbative} aspects of the $q$-deformed 
Schr\"odinger equation in the above framework.
The perturbative expansion of the $q$-deformed Hamiltonian
possesses a complex structure, which amounts to some additional 
momentum-dependent interaction 
\citer{Hebecker:1994eb,Fichtmuller:1996dt}, \cite{Zhang:2000qj}.
There are two approaches to showing such $q$-deformed effects:
One includes it in the kinetic energy term, the other includes it 
in the potential. General results are presented, and as examples, 
the harmonic-oscillator system and the Morse potential are discussed 
in some detail. 

In the following, we first review the necessary background of $q$-deformed 
quantum mechanics.
In terms of $q$-deformed phase space variables --- 
the position operator $X$ and the momentum operator $P$, 
the following $q$-deformed Heisenberg algebra has been
developed \cite{Hebecker:1994eb,Fichtmuller:1996dt}:
\begin{equation}
\label{Eq:q-algebra}
q^{1/2}XP-q^{-1/2}PX=iU, \qquad    
UX=q^{-1}XU, \qquad
UP=qPU,
\end{equation}
where $X$ and $P$ are hermitian and $U$ is unitary:
$X^{\dagger}=X$, $P^{\dagger}=P$, $U^{\dagger}=U^{-1}$.
Compared to the Heisenberg algebra the operator $U$ is a new member, 
called the scaling operator. 
The necessity of introducing the operator $U$ is as follows. 

The algebra (\ref{Eq:q-algebra}) is based on the definition of 
the hermitian momentum operator $P$. 
However, if $X$ is assumed to be a hermitian operator in a Hilbert space
the  $q$-deformed derivative \cite{Fichtmuller:1996dt,Wess:1991vh}
\begin{equation}
\label{Eq: q-derivative}
\partial_X X=1+qX\partial_X,
\end{equation}
which codes the non-commutativity of space, shows that the usual 
quantization rule $P\to -i\partial_X$ does not yield a hermitian 
momentum operator. Ref.~\cite{Fichtmuller:1996dt} showed that a hermitian 
momentum operator 
$P$ is related to $\partial_X$ and $X$ in a nonlinear way by introducing 
a scaling operator $U$
\begin{equation}
\label{Eq:scaling}
U^{-1}\equiv q^{1/2}[1+(q-1)X\partial_X], \qquad
\bar\partial_X\equiv -q^{-1/2}U\partial_X, \qquad
P\equiv -\frac{i}{2}(\partial_X-\bar\partial_X),
\end{equation}
where $\bar\partial_X$ is the conjugate of $\partial_X$. 
The operator $U$ is introduced in the definition of the hermitian momentum, 
thus it closely relates to properties of the dynamics and plays an 
essential role in $q$-deformed quantum mechanics. The nontrivial properties 
of  $U$ imply that the algebra (\ref{Eq:q-algebra}) has a richer structure
than the Heisenberg commutation relation. 
In (\ref{Eq:q-algebra}) the parameter $q$ is a fixed real number. 
It is important to make distinctions for different realizations of 
the $q$-algebra by different ranges of $q$ values \citer{Zachos,Solomon}. 
Following Refs.~\cite{Hebecker:1994eb,Fichtmuller:1996dt} we 
only consider the case $q>1$ in this paper.
In the limit $q\to 1^+$ the scaling operator $U$ reduces 
to the unit operator, thus the algebra (\ref{Eq:q-algebra}) reduces to 
the Heisenberg commutation relation. 

Such defined hermitian momentum $P$ leads to $q$-deformation effects, 
which are exhibited by the dynamical equation. 
Eq. (\ref{Eq:scaling}) shows that the  momentum $P$ depends  non-linearly
on $X$ and $\partial_X$. Thus the $q$-deformed 
Schr\"odinger equation is difficult to treat. 
In this paper we demonstrate its perturbative aspects.

The $q$-deformed phase space variables $X$, $P$ and 
the scaling operator $U$ can be realized in terms of undeformed variables 
$\hat x$, $\hat p$ of the ordinary quantum mechanics, 
where $\hat x$, $\hat p$ satisfy:
$[ \hat x, \hat p ]=i$, $\hat x=\hat x^{\dagger}$, 
 $\hat p=\hat p^ {\dagger}$.
The variables $X$, $P$ and 
the scaling operator $U$ are related to $\hat x$, $\hat p$ by
\cite{Fichtmuller:1996dt}:
\begin{equation}
\label{Eq:P-p}
X= \frac{[\hat z+\half]}{\hat z+\half}\hat x,  \qquad
P=\hat p, \qquad
U= q^{\hat z}, 
\end{equation}
where $\hat z=-\frac{i}{2}(\hat x\hat p+\hat p\hat x)$ and
$[A]$ is the $q$-deformation of $A$, defined by 
$[A]=(q^A-q^{-A})/(q-q^{-1})$. 
Using (\ref{Eq:P-p}) it is easy to check that $X$, $P$ and $U$ satisfy 
(\ref{Eq:q-algebra}). 

From (\ref{Eq:P-p}) it follows that $X$ is represented as a function of 
$\hat x$ and $\hat p$ (note that $\hat z+\half=-i\hat x \hat p$):
\begin{equation}
\label{Eq:X-variable}
X=i(q-q^{-1})^{-1} \bigl( q^{(\hat z+1/2)} -
q^{-(\hat z+1/2)}\bigr)\hat p^{-1}. 
\end{equation}
Using (\ref{Eq:X-variable}) it is convenient to discuss the perturbative 
expansion of $X$. Let $q=e^{f}=1+f$, with $0<f\ll1$. To the order $f^2$, 
   $X$ reduces to 
\begin{equation}
\label{Eq:X-perturbative}
X=\hat x  + f^2 g(\hat x, \hat p), \qquad
g(\hat x,\hat p)=-\frac{1}{6}(1+
\hat x  \hat p  \hat x  \hat p )\hat x.
\end{equation} 

The $q$-deformed phase space ($X$, $P$) governed by the $q$-algebra 
(\ref{Eq:q-algebra}) is a $q$-deformation of the ordinary quantum mechanics 
phase space ($\hat x$, $\hat p$), thus all machinery of the ordinary 
quantum mechanics can be applied to the $q$-deformed quantum mechanics. 
By analogy, dynamical equations of the quantum system are the same 
for the undeformed phase space variables $\hat x$ and $\hat p$ and 
for the $q$-deformed  phase space variables $X$ and $P$. 
Thus the starting point for establishing perturbative calculations 
of the $q$-deformed Schr\"odinger equation is as follows: 
first one uses $q$-deformed  phase space variables $X$ and $P$ 
to write down the Hamiltonian of the system, then one uses (\ref{Eq:P-p}) 
to express $X$ and $P$ by the undeformed phase space variables 
$\hat x$ and $\hat p$.                           

The $q$-deformed Hamiltonian with potential $V(X)$ is
\begin{equation}
\label{Eq:q-hamiltonian}
H(X,P)=\frac{1}{2\mu}P^{2}+V(X).
\end{equation}
For regular potentials $V(X)$, which are singularity free, to the order $f^2$ 
of the perturbative expansion, such potentials can be expressed by the 
undeformed variables $\hat x$ and $\hat p$ as
\begin{equation}
\label{Eq:q-potential}
V(X)=V(\hat x) +\hat H^{(q)}_I(\hat x,\hat p),
\end{equation}
with the perturbation
\begin{equation}
\label{Eq:H-q}
\hat H^{(q)}_I(\hat x,\hat p)=f^2\sum_{k=1}^\infty \frac{V^{(k)}(0)}{k!}
\biggl( \sum_{i=0}^{k-1}\hat x^{(k-1)-i}g(\hat x,\hat p)\hat x^{i}\biggr), 
\end{equation}
where $V^{(k)}(0)$ is the $k$-th derivative of $V(x)$ at $x=0$ 
($x$ is the spectrum of $\hat x$).
In (\ref{Eq:H-q}) the ordering between the non-commutative quantities
$ \hat x$ and $g(\hat x,\hat p)$  is  carefully considered. 
Substituting for $g(\hat x,\hat p)$ and summing over $i$,
the above result can be expressed as
\begin{equation}
\label{Eq:H-q-sumk}
\hat H^{(q)}_I(\hat x,\hat p)
=\frac{f^2}{6}\sum_{k=1}^\infty \frac{V^{(k)}(0)}{k!}\hat x^k
\biggl(k \hat x^2 \partial^2_{\hat x} +  k(k+2) \hat x \partial_{\hat x}
+\frac{1}{6}k(k-1)(2k+5)\biggr).
\end{equation}
The remaining sum over $k$ can be performed in terms of derivatives
of the potential:
\begin{equation}
\label{Eq:H-q-summed}
\hat H^{(q)}_I(\hat x,\hat p)
=\frac{f^2\hat x^2}{6}
\bigl\{ \hat x V'(\hat x) \partial^2_{\hat x} 
+ \bigl[\hat x V^{''}(\hat x) +3 V^{'}(\hat x)\bigr]\partial_{\hat x}
+{\textstyle\frac{1}{3}}\hat x V^{'''}(\hat x) 
+{\textstyle\frac{3}{2}} V^{''}(\hat x)\bigr\}.
\end{equation}

For potentials with singular term $X^{-k}$, $(k=1,2,3,\ldots)$, 
we use the following operator equation to treat the perturbation expansion:
\begin{align}                          
\label{Eq:singular}  
\frac{1}{A+B}=\frac{1}{A}-\frac{1}{A}B\frac{1}{A}+
\frac{1}{A}B\frac{1}{A}B\frac{1}{A} 
-\frac{1}{A}B\frac{1}{A}B\frac{1}{A}B\frac{1}{A}+\cdots,
\end{align}
where the norms of the operators $A$ and $B$ satisfy $\|B\|<\|A\|$.
Thus to the order $f^2$ the perturbative expansion of $1/X$ reads:
\begin{equation}                          
\label{Eq:coulomb}  
\frac{1}{X}=\frac{1}{\hat x}-f^2 \frac{1}{\hat x}\,
g(\hat x,\hat p)\, \frac{1}{\hat x}.
\end{equation}

For the energy shift, in the state $|n\rangle$,
corresponding to Eq.~(\ref{Eq:H-q-summed}),
we may integrate by parts, and obtain
\begin{equation}
\label{Eq:H-q-V}
\Delta \hat E^{(q)}_n
=-\frac{f^2}{36}
\int_{-\infty}^\infty dx\, \left\{ \psi^{(0)*}_n(x)
V(x) [2 x^3 \partial^3_{x} 
+ 9 x^2 \partial^2_{x} -3]\psi^{(0)}_n(x)+\text{h.c.}\right\},
\end{equation}
where $\psi^{(0)}_n$ is the unperturbed wave function.
One can use the Schr\"odinger equation
and rewrite this as
\begin{equation}
\label{Eq:Delta-E-hat}
\Delta \hat E^{(q)}_n
=\frac{f^2}{6}\,
\int_{-\infty}^\infty dx\, \psi^{(0)*}_n(x)
\bigl(V(x) \{1-4\mu x^2 [V(x) -E]\}
-{\textstyle\frac{2}{3}}\mu E x^3 V'(x)\bigr)\psi^{(0)}_n(x),
\end{equation}
where $E$ is the unperturbed energy.

There is another set of variables $\tilde x$ and $\tilde p$ 
of an undeformed algebra, which are obtained by a canonical transformation 
of $\hat x$ and $\hat p$ \cite{Fichtmuller:1996dt}:
\begin{equation}
\label{Eq:tilde}
\tilde x=\hat x F^{-1}(\hat z), \qquad \tilde p= F(\hat z)\hat p,
\end{equation}
where (note that $\hat z-\half=-i\hat p \hat x$)
\begin{equation}
\label{Eq:F(z)}
F^{-1}(\hat z)= \frac{[\hat z-\half]}{\hat z-\half}, \qquad
\end{equation}
Such defined variables $\tilde x$ and $\tilde p$ also satisfy the undeformed 
algebra: $[ \tilde x, \tilde p ]=i$, and 
$\tilde x=\tilde x^{\dagger}$,\quad$\tilde p=\tilde p^{ \dagger}$.
Thus $\tilde p=-i\partial_{\tilde x}$.
The $q$-deformed variables $X$, $P$ and 
the scaling operator $U$ are related to  $\tilde  x$ and $\tilde p$ as follows:
\begin{equation}
\label{Eq:X-x}
X=\tilde x, \qquad P=F^{-1}(\tilde z) \tilde p, \qquad
U= q^{\tilde z}, 
\end{equation}
where $\tilde z=-\frac{i}{2}(\tilde x\tilde p + \tilde p\tilde x)$;
and with $F^{-1}(\tilde z)$ defined by Eq.~(\ref{Eq:F(z)}) for the variables  
($\tilde  x$, $\tilde p$).
From Eqs.~(\ref{Eq:tilde})--(\ref{Eq:X-x}) it follows that $X$,
$P$ and $U$ also satisfy (\ref{Eq:q-algebra}), and Eq. (\ref{Eq:X-x})  is 
equivalent to Eq. (\ref{Eq:P-p}).

Using (\ref{Eq:X-x}) to the order $f^2$ the perturbative expansions of $P$
and the kinetic energy $P^{2}/(2\mu)$ read 
\begin{equation}
\label{Eq:P-perturbative}
P=\tilde p  + f^2 h(\tilde x ,\tilde p),\qquad
h(\tilde  x,\tilde p)
=-\frac{1}{6}(1+
\tilde  p\tilde x\tilde  p\tilde x )\tilde p,
\end{equation} 
and
\begin{equation}
\label{Eq:Ek-perturbative}
\frac{1}{2\mu}P^{2}=\frac{1}{2\mu}\tilde p^{2}+
\tilde H^{(q)}_I(\tilde x,\tilde p), 
\end{equation} 
with
\begin{eqnarray}
\label{Eq:tilde-Hq}
\tilde H^{(q)}_I(\tilde x,\tilde p)&=& \frac{1}{2\mu}f^2
\bigl[\tilde p\, h(\tilde  x,\tilde p)
+h(\tilde  x,\tilde p)\, \tilde p\bigr] 
\nonumber\\
&=& -\frac{1}{12\mu}f^2 \bigl[ 2\tilde x^2 \partial_{\tilde x}^4+
8\tilde x  \partial_{\tilde x}^3+
3\partial_{\tilde x}^2 \bigr]
\end{eqnarray}
Eqs.~(\ref{Eq:Ek-perturbative}) and (\ref{Eq:tilde-Hq}) show that in the 
($\tilde x$,$ $ $\tilde p$) system the perturbative contribution comes 
from the  kinetic-energy term.

Similar to Eqs.~(\ref{Eq:H-q-V})--(\ref{Eq:Delta-E-hat})
(using the Schr\"odinger equation and integrating by parts), 
one can write the energy shift 
corresponding to Eq.~(\ref{Eq:tilde-Hq}) as
\begin{equation}
\label{Eq:Delta-E-tilde}
\Delta\tilde E^{(q)}_n
=\frac{f^2}{6}
\int_{-\infty}^\infty dx\, \psi^{(0)*}_n(x)
[V(x)-E]\{1-4\mu x^2[V(x)-E]\}\psi^{(0)}_n(x).
\end{equation}

The two expressions for the energy shift,
Eqs.~(\ref{Eq:Delta-E-hat}) and (\ref{Eq:Delta-E-tilde}) are in fact
equal, since the difference is given by
\begin{equation}
\frac{f^2}{6}\,E
\int_{-\infty}^\infty dx\, \psi^{(0)*}_n(x)
\bigl\{
1-4\mu x^2[V(x)-E]
-{\textstyle\frac{2}{3}} x^3\mu V'(x)\bigr\}\psi^{(0)}_n(x)=0.
\end{equation}
From this last form, Eq.~(\ref{Eq:Delta-E-tilde}),
it is easy to see that the energy shift is negative
since $\langle n|V|n\rangle < E$. Thus,
\begin{equation}
\Delta E^{(q)}_n < 0.
\end{equation}

As a first application we consider the  $q$-deformed ``harmonic'' system 
described by the Hamiltonian
\begin{equation}                          
\label{Eq:qd-harmonic}
H(X,P)=\frac{1}{2\mu}P^2+\frac{1}{2}\mu\omega^2 X^2,
\end{equation}

First we calculate $\Delta {\tilde E^{(q)}_n}$ in the $(\tilde x,\tilde p)$ 
system. From Eq.~(\ref{Eq:tilde-Hq}) or (\ref{Eq:Delta-E-tilde}) 
it follows that the shifts of the energy levels are
\begin{equation}
\label{Eq:Delta-E}
\Delta {\tilde E}^{(q)}_n
=-\frac{f^2\omega}{48}\left(4n^3+6n^2+20n+9\right).
\end{equation}

In the $(\hat x,\hat p)$ system the only non-zero term in (\ref{Eq:H-q})
is $V^{(2)}(0)=\mu\omega^2$, thus (\ref{Eq:H-q}) reduces to:
\begin{eqnarray}
\label{Eq:harmonic}
\hat H^{(q)}_I(\hat x,\hat p))
&=&-\frac{1}{12}f^2\mu\omega^2\bigl[\hat x(1+\hat x\hat p\hat x\hat p)
\hat x+(1+\hat x\hat p\hat x\hat p)\hat x^2\bigr]
\nonumber \\
&=&\frac{1}{12}f^2\mu\omega^2
\bigl[ 2 \hat x^4 \partial_{\hat x}^2+8\hat x^3\partial_{\hat x}
+3\hat x^2\bigr].
\end{eqnarray}
The corresponding energy shift, which can also be obtained
from Eq.~(\ref{Eq:Delta-E-hat}), is easily seen to be identical
to that of Eq.~(\ref{Eq:Delta-E}).

As noted above, the shift in Eq.~(\ref{Eq:Delta-E}) is negative, 
and it increases with $n$, 
leading eventually to a breakdown of perturbation theory for 
$n\sim(12/f^2)^{1/3}$.
The tendency exhibited by Eq.~(\ref{Eq:Delta-E}) agrees with the observation
that for the $q$-deformed harmonic oscillator the spectrum has
an upper bound \cite{Lorek:1997xs}.

In the limiting case $q\to1^+$ we have $H(X,P)\to H_{\rm un}(\hat x,\hat p)=
\frac{1}{2\mu}\hat p^{2}+\frac{1}{2}\mu\omega^2\hat x^2$. 
Only in this sense $H(X,P)$ defined in Eq.  (\ref{Eq:qd-harmonic}) is called 
the  $q$-deformed ``harmonic'' system. 

As another example, we study the Morse potential \cite{Morse}
in its ``supersymmetric'' form \cite{Dutt:1988va},
where the ground state energy vanishes.
It is given by the potential
\begin{equation}
V(x)=A^2+B^2e^{-2\alpha x}
-2B\left(A+\frac{\alpha}{2\sqrt{2\mu}}\right)e^{-\alpha x}.
\end{equation}
The corresponding energy shift can be obtained from either
Eq.~(\ref{Eq:Delta-E-hat}) or Eq.~(\ref{Eq:Delta-E-tilde}),
the result is shown in Fig.~1 for $\alpha=1$, $\mu=1$, and some
range of $A$ and $B$. 
For the harmonic oscillator, we saw that the shift increased
in magnitude with the unperturbed energy. 
This is not the case for the Morse potential, 
where the shift may increase or decrease with
the unperturbed energy, depending on the parameters.

\begin{figure}[htb]
\refstepcounter{figure}
\label{Fig:twodim}
\addtocounter{figure}{-1}
\begin{center}
\setlength{\unitlength}{1cm}
\begin{picture}(8.0,8.0)
\put(0,-0.3)
{\mbox{\epsfysize=8.5cm\epsffile{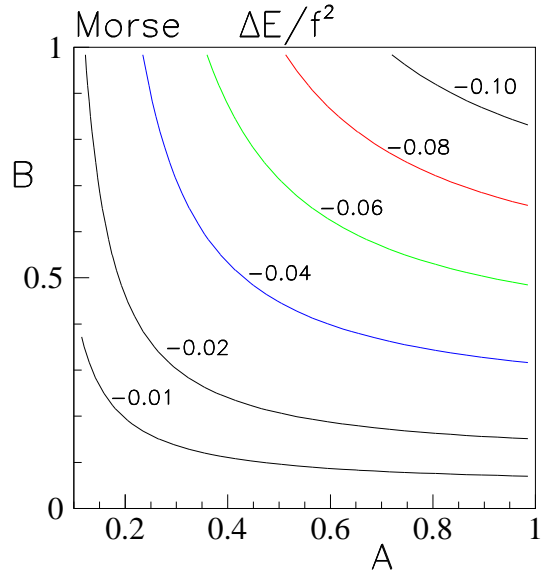}}}
\end{picture}
\caption{Energy shift for the Morse potential, 
$\Delta E/f^2$, {\it vs.} $A$ and $B$, for $\alpha=1$.}
\end{center}
\end{figure}

It should be emphasized again that $\tilde H^{(q)}_I(\tilde x,\tilde p)$ 
originates from the kinetic term, whereas $\hat H^{(q)}_I(\hat x,\hat p)$ 
originates from the potential. 
At the level of operators, these two Hamiltonians are different.
However, they differ only by a quantity whose expectation value vanishes.

At short distances, where $q$-deformation might be relevant,
one also expects quantum mechanics to break down and have to be replaced
by some kind of field theory.
Some progress is being made in this area \cite{Chaichian:2000ba}.
In a more realistic theory along such lines, some features
of $q$-deformed quantum mechanics may survive.
It is therefore hoped that studies of $q$-deformed dynamics at the level
of quantum mechanics will give some clue for the further development.
\medskip

This work has been supported by 
DAAD--K.C.Wong Fellowship (Germany) and the
Research Council of Norway.
JZZ would like to thank Prof.\ H.J.W.~M\"uller-Kirsten for helpful
discussions. His work has also been supported by the National Natural Science 
Foundation of China under the grant number 10074014 and by the Shanghai 
Education Development Foundation. 


\end{document}